\documentclass[preprint,aps,nofootinbib,superscriptaddress,11pt]{revtex4-1}
\usepackage{amssymb,mathrsfs,physics,amsthm,amssymb,amsfonts,amsmath}

\usepackage{import}

\usepackage{hyperref}
\hypersetup{
    plainpages=false,       
    unicode=false,          
    pdftoolbar=true,        
    pdfmenubar=true,        
    pdffitwindow=false,     
    pdfstartview={FitH},    
    pdfnewwindow=true,      
    colorlinks=true,        
    linkcolor=Blue,         
    citecolor=Blue,        
    filecolor=magenta,      
    urlcolor=blue           
}
\usepackage[usenames, dvipsnames]{xcolor}

\newcommand\id{\leavevmode\hbox{\small1\kern-3.3pt\normalsize1}}

\usepackage{enumitem}

\usepackage[normalem]{ulem}

\theoremstyle{definition}
\newtheorem{theorem}{Theorem}

\usepackage{chngcntr}
\usepackage{apptools}
\AtAppendix{\counterwithin{theorem}{section}}

\usepackage{upgreek}

\usepackage{graphicx}
\graphicspath{ {images/} }

\usepackage{caption}

\usepackage{tabularx}

\usepackage{bbm}

\usepackage{tikz}
\usetikzlibrary{arrows}

\newcommand{\cps}{\mathrel{\ooalign{$\nearrow$\cr \kern-0pt$\nwarrow$}}}

\usepackage{CJKutf8}

\usepackage{versions}
\excludeversion{cmt}

\newcommand{\gs}{{g^\circ}}
\newcommand{\gv}{{g^\bullet}}

\newcommand{\asi}{{a^{\circ}_1}}
\newcommand{\aso}{{a^{\circ}_2}}
\newcommand{\avi}{{a^{\bullet}_1}}
\newcommand{\avo}{{a^{\bullet}_2}}
\newcommand{\bsi}{{b^{\circ}_1}}
\newcommand{\bso}{{b^{\circ}_2}}
\newcommand{\bvi}{{b^{\bullet}_1}}
\newcommand{\bvo}{{b^{\bullet}_2}}

\newcommand{\As}{{A^{\circ}}}
\newcommand{\Bs}{{B^{\circ}}}
\newcommand{\Es}{{E^{\circ}}}

\newcommand{\Av}{{A^{\bullet}}}
\newcommand{\Bv}{{B^{\bullet}}}
\newcommand{\Ev}{{E^{\bullet}}}

\begin{document}

\title{Analogue of Null Geodesic in Quantum Spacetime}
\begin{CJK*}{UTF8}{gbsn}
\author{Ding Jia (贾丁)}
\email{ding.jia@uwaterloo.ca}
\affiliation{Department of Applied Mathematics, University of Waterloo, Waterloo, Ontario, N2L 3G1, Canada}
\affiliation{Perimeter Institute for Theoretical Physics, Waterloo, Ontario, N2L 2Y5, Canada}

\begin{abstract}
In classical spacetime null geodesics give information on where free massless particles travel. In quantum spacetime where quantum indefiniteness renders spacetime fuzzy null geodesics as sharp trajectories cannot be retained. We propose a ``tendency postulate'' that gives information on where free massless objects tend to travel based exclusively on meaningful notions in quantum spacetime. It says that free massless objects tend to trace large quantum causal fluctuations. We implement the tendency postulate in concrete models of quantum spacetime that describe causal fluctuation. The tendency postulate supports the suggestion that information can leak out of black holes in a quantum spacetime without introducing superluminal signalling.
\end{abstract}
 
\maketitle
\end{CJK*}

\begin{cmt}
causal fluctuation rather than gravitational flucutuation
\end{cmt}

\section{Introduction}\label{sec:i}

In studying classical gravity we rely much on the notion of null geodesic, with various other key notions such as the null geodesic congruence, the light-cone, the light-sheet, and the various horizons (the event horizon, the apparent horizon, the Killing horizon etc.) defined through it. However, the null geodesic as a sharp test particle trajectory cannot be retained in quantum spacetime where quantum indefiniteness renders spacetime fuzzy. An analogue notion for quantum spacetime is in need.

In this paper we propose a ``tendency postulate'' which says that \textit{in quantum spacetime free massless objects tend to trace large quantum causal fluctuations}. Just as null geodesics offer information on where free massless particles travel in classical spacetime, the tendency postulate offers information on where massless objects travel in quantum spacetime. In this sense, the tendency postulate provides a quantum spacetime analogue of null geodesics.

The postulate is motivated by the expectation that quantum gravitational fluctuations are large at small distances. Crucially, in keeping with the spirit of relativity the smallness of the distance is measured in the invariant absolute value of the proper distance rather than spatial or temporal distances. This means that large quantum gravitational fluctuations are expected for close-to-null separations, which is what a small absolute value of proper distance means. In classical spacetime it implies that null trajectories tend to accompany large quantum gravitational fluctuations. The tendency postulate simply reformulates the above idea using terms meaningful in quantum spacetime, without referring to trajectories or distances.

The tendency postulate refers specifically to ``quantum causal fluctuations'', the causal part of quantum gravitational fluctuations, even though the classical spacetime metric contains a conformal part and a causal part \cite{hawking1973large, hawking1976new, malament1977class}. In classical spacetime the causal part determines the trajectory of free massless test particles completely. We extrapolate this feature to quantum spacetime and assume that the tendency postulate, which gives analogous information on where massless objects travel, needs to only refer to the causal part of the quantum gravitational fluctuations. A very intuitive support is that in classical spacetime light-like separation lies critically between time-like separation and space-like separation -- precisely null geodesics are most susceptible to perturbations of the causal structure. This strongly suggests a direct connection between large quantum causal fluctuations and light-like separations, or to use terms meaningful in quantum spacetime, high tendencies of free massless object travellings.

The tendency postulate refers to ``massless'' objects rather than massless ``particles''. This has to do with the subtleties of particle as a fundamental concept of physics. A lesson from quantum field theory is that due to effects such as Unruh's, the notion of particle is observer-dependent \cite{birrell1984quantum}. Furthermore, there is the view (e.g. \cite{wald1994quantum}) that fields are more fundamental concepts than particles, which are derived as excitations of the fields. It is unclear how it will turn out for the notion of particle in quantum spacetime. To free ourselves from completely from the subtleties related to whether the notion of particle should show up in a statement about fundamental properties of quantum spacetime such as the tendency postulate, we simply refrain from referring to massless particles. Instead, we talk about massless objects, which we do not attempt to rigorously define but resort to intuition to denote massless things that carry energy and transmit information.

The tendency postulate is a statement about what massless objects \textit{tend} to do, rather than what they always do. This is to account for the probabilistic nature of a quantum spacetime. In this respect the tendency postulate is analogous to the second law of thermodynamics, which is a statistical law that can be violated with finite probabilities. The tendency postulate allows massless object to not trace large quantum causal fluctuations with small probability, though most of the time the tendency is obeyed.

The expression ``tend to trace'' in the tendency postulate means the following. Suppose the party $A$ sends out some massless objects and several other parties $B_1, B_2,\cdots, B_n$ can possibly receive them. For each $B_i$, one can infer (as specified below) from the description of quantum spacetime the strength of the quantum causal fluctuation associated with the pair of parties $A$ and $B_i$. The tendency postulate says that if the massless objects freely propagate, with large probability they will arrive at those $B_i$ for which the strength of the quantum causal fluctuation associated with $A$ and $B_i$ is large. Here ``tend'' refers to the large probability, and ``trace'' refers to picking and following the large quantum causal fluctuations.

We recognize three uses of the tendency postulate. It can firstly be used to develop further concepts for quantum spacetime which used to be based on null geodesics for classical spacetime. These include the null geodesic congruence, the light-cone, the light-sheet, and the various horizons. Moreover, the tendency postulate can be used to pick out typical physical states among all possible ones allowed by the frameworks. For example, QFT admits a large family of states (In algebraic QFT all normalized linear forms over the algebra of local observables are regarded as states \cite{haag1964algebraic}.), and criteria such as the Hadamard condition \cite{kay1991theorems}, the wavefront set condition \cite{radzikowski1992hadamard, radzikowski1996micro}, and the principle of local definiteness \cite{haag1984quantum} were proposed to select the physical ones. The tendency postulate can similarly be used as a criterion for physically reasonable states to obey. Note that, as will be further discussed later in Section \ref{sec:tp}, the tendency postulate as a statement about the probabilistic quantum spacetime characterizes typical models but cannot rule out atypical models. Finally, the tendency postulate can be used to come up with particular examples that are typical. Suppose one wants to use concrete models of quantum spacetime to test some ideas. A good strategy is to construct models using the tendency postulate as a guide.

The main focus of the paper is to demonstrate how the tendency postulate can be implemented concretely in models of quantum spacetime even without having a complete theory of quantum gravity. This is possible thanks to progresses made in the previous two decades in modelling indefinite causal structure in operational formulations of quantum theory and more general probabilistic theories, starting with the seminal work of Hardy \cite{hardy2005probability}. In models of quantum spacetime with indefinite causal structure it is possible to determine where large quantum causal fluctuation takes place. 


Broadly speaking, studying quantum causality in operational theories is emerging as a new approach to attack questions in quantum gravity. The new approach is a ``top-down'' one in H{\"o}hn's characterization \cite{hohn2017reflections} and complements the ``bottom-up approaches'' such as string theory, loop quantum gravity and causal set theory. The bottom-up approaches investigate quantum gravity at the most fundamental level building on postulates about the basic degrees of freedom of quantum spacetime. The top-down approaches on the other hand study those aspects of quantum gravity that are independent of what exactly the basic degrees of freedom are for quantum spacetime. While an ultimate understanding of quantum gravity will probably need contributions from both kinds of approaches, the top-down ones allow us to overcome a major difficulty in the subject of quantum gravity -- the lack of empirical data to judge among the different assumptions about the basic degrees of freedom of quantum spacetime. Very importantly, the study of this paper following a top-down approach should be compatible with any theory of quantum gravity where the causal structure of spacetime quantum fluctuates. Any such theory should admit an operational level description of the correlated probabilities in the presence of causal fluctuations, which is essentially the only ingredient that is needed to state and implement the tendency postulate.

We are hopeful that through studying quantum causality in a top-down approach, some general lessons independent of what spacetime is made of at the basic level can be learned. As an attempt to make progress along this direction, we show how the tendency postulate supports the view that information can leak out of black holes without introducing superluminal signalling when quantum causal fluctuation is brought into consideration.

The plan of the paper is as follows. In Sections \ref{sec:fc} and \ref{sec:model} we show how to model quantum causal fluctuation in operational formulations of quantum theory. In Section \ref{sec:sec} we introduce the notion of massless and massive sectors in order to implement the tendency postulate in Section \ref{sec:tp}. In Section \ref{sec:ss} we discuss information leakage from black holes, along with the question of how to set up a quantum-classical spacetime correspondence and why information leakage does not lead to superluminal signalling. We conclude with some general discussions and outlooks in Section \ref{sec:do}.

\section{Framework and conventions}\label{sec:fc}

Since the seminal work of Hardy \cite{hardy2005probability}, several theories appeared that can be used to describe indefinite causal structure (e.g. \cite{hardy2007towards, hardy2012operator, chiribella2012perfect, chiribella2013quantum, oreshkov2012quantum, oreshkov2015operational, maclean2017quantum, oeckl2016local}, see \cite{brukner2014quantum} for an overview of the field). The theories differ in levels of generality and some fine details, but they are commonly based in the operational probabilistic setting. One primary goal of this paper is to demonstrate how the tendency postulate can be implemented concretely in a model of quantum spacetime with indefinite causal structure. We choose to work with Oreshkov, Costa and Brukner's process matrix framework \cite{oreshkov2012quantum}, which supports some models of quantum spacetime that are exactly solvable for the causality measure relevant to our study. We stress that the tendency postulate is a model independent statement about quantum spacetime and can be implemented for other models.

The main idea of the process matrix framework \cite{oreshkov2012quantum, araujo2015witnessing, oreshkov2016causal} is to assume that ordinary quantum theory with fixed causal structure holds locally, while globally the causal structure can be indefinite. The framework studies process matrices, which encode the global causal structure among the local parties.

Local parties, or parties in short, are localized spacetime regions where agents (including humans and nature itself) conduct their operations. We do not associated the expression of ``localized spacetime region'' with any specific mathematical structure such as a differentiable manifold in general relativity. Rather, the expression ``localized spacetime region'' simply captures the intuition that an agent's action takes place within some finite spatial region and lasts for some finite temporal duration. We denote the parties by capital letters $A, B, \cdots$. Each such party must receive some information carried by a physical system from the past and send out some information carried by a physical system to the future. Therefore we associate a party $A$ with an input system $a_1$ and an output system $a_2$ (each system $x$ is associated with a Hilbert space $\mathcal{H}^{x}$). In the probabilistic theory setting, a general operation is accompanied by gaining some classical information in terms of a classical outcome $i$ within a possible set of outcomes $\mathcal{I}$. The special case of no gaining of classical information corresponds to a singleton $\mathcal{I}$. 

A process is simply a general object that allow one to derive probability of correlated classical outcomes of local parties. For example, a bipartite state for parties $A$ and $B$ is a process that allows one to deduce correlated measurement probabilities for the pair of parties. A channel from $A$ to $B$ is also a process that allows one to deduce correlated preparation (imagine $A$ prepares a state from a set of states with certain probabilities) and measurement probabilities for the pair of parties. More generally, a process is required to obey three rather weak conditions called local quantum mechanics, non-contextuality, and extendibility \cite{oreshkov2016causal}. Local quantum mechanics says that ordinary quantum theory with fixed causal structure holds in local parties. Non-contextuality says that the joint outcome probability of local parties is non-contextual, i.e., equivalent local operations lead to the same probabilities. Extendibility says that the operations of local parties can be extended to act on an arbitrary joint ancilla quantum state (see \cite{oreshkov2016causal} for details). For a bipartite state or a channel, the correlated outcome statistics imply some definite causal relations between the parties. For a general process obeying the three conditions above, however, it is possible that one cannot assign a definite causal relation to $A$ and $B$ to reproduce the statistics. Therefore a general process incorporates indefinite causal structure. Although it is not clear whether all processes obeying the mathematical conditions above can be created in laboratory or observed in nature, there have been claims of laboratory detection of certain processes with indefinite causal structure at the classical or quantum level \cite{procopio2015experimental, maclean2017quantum, rubino2017experimental}. In this paper we focus on studying process matrices that we believe are induced by the nature's quantum gravitational effects.

So far we talked about processes as some abstract object that allows deduction of correlated statistics. Now we find the concrete representation of processes as ``process matrices'' (we assume all systems are finite-dimensional for simplicity). The strategy is to first represent local operations as matrices, then consider processes as maps from local operations to probabilities, and use this duality structure to represent processes as matrices.

It is standard in quantum theory to represent local operations that come with classical outcomes as quantum instruments \cite{davies1970operational, nielsen2000quantum}. A quantum instrument is a set of completely positive (CP) trace-non-increasing maps $\{N_i\}_{i\in \mathcal{I}}$ that sum up to a CP trace-preserving map (channel) $N=\sum_{i\in \mathcal{I}} N_i$. Let $L(\mathcal{H})$ stand for the linear operators on Hilbert space $\mathcal{H}$, and $\abs{x}$ stand for the dimension of the Hilbert space $\mathcal{H}^x$. Each CP map $M:L(\mathcal{H}^{a_1})\rightarrow L(\mathcal{H}^{a_2})$ can be represented isomorphically by its ``Choi operator'' defined as follows\footnote{The $\abs{a_1a_2}$ is absorbed into the Choi operators for the local instrument here so that the process matrices have trace one. This is merely conventional and one can choose to absorb the factor into the process matrices without changing the observational outcome probabilities.} \cite{choi1975completely}:
\begin{align}
\hat{M}=\abs{a_1a_2} M\otimes \id (\ketbra{\Upphi}{\Upphi})\in L(\mathcal{H}^{a_2}\otimes\mathcal{H}^{a_1}),\label{eq:cs}
\end{align}
where $\id$ is the identity channel on system $a_1$, $\ket{\Upphi}=\sum_i \abs{a_1}^{-1/2}\ket{ii}\in \mathcal{H}^{a_1}\otimes \mathcal{H}^{a_1}$ is a normalized maximally entangled state in a canonical basis on two copies of system $a_1$.

A correlated classical outcome in parties $A,B,\cdots,C$ is labeled by $(i,j,\cdots,k)$ and is associated with a set of instrument maps $M_i,N_j,\cdots,L_k$ with operators $\hat{M}_i,\hat{N}_j,\cdots,\hat{L}_k$. A process assigns to each such outcome and such operator a probability, so can be viewed as a map from $L(\mathcal{H}^{a_2}\otimes\mathcal{H}^{a_1})\otimes L(\mathcal{H}^{b_2}\otimes\mathcal{H}^{b_1})\otimes \cdots\otimes L(\mathcal{H}^{c_2}\otimes\mathcal{H}^{c_1})$ to the real numbers assuming linearity. By standard functional analysis (see \cite{oreshkov2012quantum} for details), the process maps can be represented as linear operators $W\in L(\mathcal{H})$, where $\mathcal{H}:=\mathcal{H}^{a_1}\otimes\mathcal{H}^{a_2}\otimes \mathcal{H}^{b_1}\otimes\mathcal{H}^{b_2} \otimes \cdots \otimes \mathcal{H}^{c_1}\otimes\mathcal{H}^{c_2}$, and the generalized Born rule says that the probability of observing the joint outcome $(i,j,\cdots,k)$ is given by
\begin{align}\label{eq:br}
P\big(M_i,N_j,\cdots, L_k\big)=\Tr[(\hat{M}_i\otimes \hat{N}_j\otimes\cdots\otimes \hat{L}_k)^T W],
\end{align}
where $T$ denotes operator transpose. The requirement that probabilities are non-negative and normalized imply that $W$ is positive semi-definite and has trace $1$ plus that $W$ lives in a linear subspace of the whole space, an additional condition which we will not use in this paper. 

We note that not only are channels and states special cases of processes, but ordinary quantum theory with definite causal structure is a subtheory within the process matrix framework. The process matrix framework is a therefore generalization of ordinary quantum theory to incorporate indefinite causal structure.

The ``tensorial'' notation of \cite{hardy2011reformulating, hardy2012operator} is convenient in distinguishing input and output systems (and in writing down composition of different elements, although we do not make much use of this convenience in this paper). We label the operations and processes with superscripts and subscripts to represent output and input systems respectively. For example, the local map $M$ inside party $A$ with input $a_1$ and output $a_2$ will be written as $M_{a_1}^{a_2}$. Similarly we write $N_{b_1}^{b_2}$ for the local map $N$ in $B$. On the other hand, the process $W$ takes the output systems of $M$ and $N$ as its own input systems (information flows out of $M$ and $N$ moves into $W$), and the input systems of $M$ and $N$ as its own output systems (information flows out of $W$ and moves into $M$ and $N$). Therefore $W$ will be written as $W^{a_1b_1}_{a_2b_2}$. The composition $W(M,N)$ of feeding $M$ and $N$ into the map $W$ to get a probability according to (\ref{eq:br}) will be written as $M_{a_1}^{a_2}W^{a_1b_1}_{a_2b_2}N_{b_1}^{b_2}\in \mathbb{R}$. The repeated indices in the super- and subscripts are composed, analogous to the Einstein convention for tensors.

In contexts where the the compositional structure is not important we may still use expressions such as $W$ without super- and subscripts. Sometimes we specify the relevant parties for a process explicitly by expressions such as $W^{AB}$. Finally, we denote the maximally mixed state as $\pi$ and the canonical maximally entangled state as $\Upphi$ (as already used above for the Choi operator). We denote the causal relations of $A$ causally precede $B$, $A$ causally succeed $B$, and $A$ causally disconnected from $B$ by $A\rightarrow B, A\leftarrow B, A - B$, respectively.


\section{Clean models and general models}\label{sec:model}

In this section we present some explicit process matrices that model causal fluctuations of quantum spacetime. The so-called ``harmonic clean models'' can be regarded as canonical examples in studying quantum causal fluctuations, just like harmonic oscillators in studying quantum mechanics.

\subsection{Clean models}

A \textit{clean model} contains a separate quantum gravitational party that induce indefinite causal structure for the other parties. Such a model is ``clean'' in the sense that the quantum gravitational degrees of freedom cleanly separate from other degrees of freedom, and that there is a basis of the quantum gravitational party whose elements each couples to a definite causal relation between $A$ and $B$. A general model (cf. Subsection \ref{subsec:gm}) may not have this property.

Let us start some simple clean models with respect to two parties as an illustration. Let $A$ and $B$ share a process matrix $W^{GAB}$ that is correlated with a quantum gravitational party $G$. Start with the ansatz that $\Tr_G W^{GAB}=\sum_{i=1}^3 p_i W_i^{AB}$, where $\sum_{i=1}^3 p_i=1$, and $W_i^{AB}$ are bipartite process matrices with the causal relations $A\rightarrow B, A\leftarrow B, A - B$ for $i=1,2,3$, respectively. This says that when no information from the quantum gravity party $G$ is available regarding $W^{ABG}$, the process matrix $W^{AB}$ that is accessible to $A$ and $B$ is described as if it was a classical mixture. In general, the $W_i^{AB}$ may be correlated with the environment so they can be ``purified'' into $\ketbra{w_i}^{ABE}$ where the environmental party $E$ only contains systems that send information into it (cf. Figure \ref{fig:3cr}). More precisely, to ``purify'' $W_i^{AB}$ we view it as a channel from its systems $a_2$ and $b_2$ that receive information to its systems $a_1$ and $b_1$ that send out information, and then find the isometric extension $\ketbra{w_i}^{ABE}$ of the channel. We further arranged the environmental party $E$ to be the same for all $i$ (by enlarging $E$ to incorporate all the individual purifying systems for each $i$, if necessary). 

The process matrix for $GAB$ is 
\begin{align}
W^{GAB}(\alpha)=&\Tr_E W^{GABE}(\alpha)=\Tr_E \ketbra{w(\alpha)}^{GABE},
\end{align}
where
\begin{align}\label{eq:cm}
\ket{w(\alpha)}^{GABE}:=&\alpha_1\ket{1}^G\ket{w_1}^{ABE}+\alpha_2\ket{2}^G\ket{w_2}^{ABE}+\alpha_3\ket{3}^G\ket{w_3}^{ABE}.
\end{align}
The second equation is the defining equation for the model. We introduced the complex 3-vector $\alpha=(\alpha_1,\alpha_2,\alpha_3)$ obeying $\abs{\alpha_i}^2=p_i$ and  $\abs{\alpha_1}^2+\abs{\alpha_2}^2+\abs{\alpha_3}^2=1$ to represent the probability amplitudes for the three causal relations. One can check that the ansatz $\Tr_G W^{GAB}=\sum_{i=1}^3 p_i W_i^{AB}$ is matched with $W_i^{AB}=\Tr_E\ketbra{w_i}^{ABE}$. In modelling causal fluctuations, the quantum gravitational party $G$ is the ``active'' source of the fluctuation, which leads $A$ and $B$ into particular causal configurations parametrized by $\alpha$.

In general, a process matrix $W^{GAB}$ is a \textit{clean model} if it is of the form
\begin{align}
W^{GAB}(\alpha)=\Tr_E W^{GABE}(\alpha)=\Tr_E \ketbra{w(\alpha)}^{GABE},
\end{align}
where
\begin{align}\label{eq:gcm}
\ket{w(\alpha)}^{GABE}:=&\sum_i^n\alpha_i\ket{i}^G\ket{w_i}^{ABE}, \quad \sum_i^n\abs{\alpha_i}^2=1, \quad \{\ket{i}\}_{i=1}^n\text{ is an ONB for }G.
\end{align}
The assumptions behind this form are that there is a quantum gravitational party $G$ with its own system separated from the systems of $A$ and $B$, that the orthonormal basis vectors $\{\ket{i}\}$ each couple to a $\ket{w_i}^{ABE}$ that generates a process matrix $\ketbra{w_i}^{ABE}$ with a definite causal relation $A\rightarrow B, A\leftarrow B$ or $A - B$, and that $\ket{w_i}^{ABE}$ in a ``purified'' from is possible in the first place.


The much studied ``quantum switch'' \cite{chiribella2013quantum}, the 2-causal-relation model of \cite{feix2017quantum} and the 3-causal-relation models of \cite{jia2017quantum} are all examples of clean models.

\subsection{Harmonic clean models}\label{subsec:scm}

\begin{figure}
    \centering
    \resizebox{1\textwidth}{.36\textwidth}{
\begin{tikzpicture}

\fill [yellow!50] (11,-4.5)  -- (8.5,4) -- (17,4) -- (14.5,-4.5) -- cycle;
\fill [green!30] (10.5,-2)  -- (9.5,4) -- (13.5,4) -- (12.5,-2) -- cycle;
\fill [purple!25] (10.5,1.5)  -- (10,4) -- (13,4) -- (12.5,1.5) -- cycle;
\draw[thick,red,dashed, fill=white]  (11.5,-2) circle (1);
\node[red] at (11.5,-2) {\huge $A$};
\draw[thick,blue,dashed,fill=white]  (11.5,1.5)circle (1);
\node[blue]  at (11.5,1.5) {\huge $B$};
\draw[thick, green] (7.5,4) -- (18,4);
\node[green] at (13,5.5) {\huge Environment $\mathsf{e}$};
\node at (12.75,-5.25) {\huge $\Psi$};
\draw[thick]  (11,-4.5) rectangle (16,-6);
\draw[-triangle 90] [thick]  (14,-4.5) to (14,4);
\draw[-triangle 90] [thick]  (11.5,-4.5) to (11.5,-3);
\draw[-triangle 90] [thick]  (11.5,2.5) to (11.5,4);
\draw[-triangle 90] [thick]  (11.5,-1) to (11.5,0.5);

\node at (12,3) {\huge $\mathsf{b_2}$};
\node at (12,0) {\huge $\mathsf{b_1}$};
\node at (11,-0.5) {\huge $\mathsf{a_2}$};
\node at (11,-3.5) {\huge $\mathsf{a_1}$};

\fill [yellow!50] (34,-4.5)  -- (31.5,4) -- (40,4) -- (37.5,-4.5) -- cycle;
\fill [green!30] (33,1.5)  -- (32.5,4) -- (35.5,4) -- (35,1.5) -- cycle;
\fill [purple!25] (36.5,1.5)  -- (36,4) -- (39,4) -- (38.5,1.5) -- cycle;
\draw[thick,red,dashed,fill=white]  (34,1.5)circle (1);
\node[red] at (34,1.5) {\huge $A$};
\draw[thick,blue,dashed,fill=white]  (37.5,1.5)circle (1);
\node[blue] at (37.5,1.5) {\huge $B$};
\draw[thick, green] (30.5,4) -- (41,4);
\node[green] at (36,5.5) {\huge Environment $\mathsf{e}$};
\node at (35.75,-5.25) {\huge $\Psi$};
\draw[thick]  (34,-4.5) rectangle (39,-6);
\draw[-triangle 90] [thick]  (37,-4.5) to (37.5,0.5);
\draw[-triangle 90] [thick]  (34.5,-4.5) to (34,0.5);
\draw[-triangle 90] [thick]  (34,2.5) to (34,4);

\draw[-triangle 90] [thick]  (37.5,2.5) to (37.5,4);
\node at (33.5,3) {\huge $\mathsf{a_2}$};

\node at (38,0) {\huge $\mathsf{b_1}$};
\node at (38,3) {\huge $\mathsf{b_2}$};
\node at (33.5,0) {\huge $\mathsf{a_1}$};

\fill [yellow!50] (22.5,-4.5)  -- (20,4) -- (28.5,4) -- (26,-4.5) -- cycle;
\fill [purple!25] (24.5,-2)  -- (23.5,4) -- (27.5,4) -- (26.5,-2) -- cycle;
\fill [green!30] (24.5,1.5)  -- (24,4) -- (27,4) -- (26.5,1.5) -- cycle;

\draw[thick,red,dashed, fill=white]  (25.5,1.5) circle (1);
\node[red] at (25.5,1.5) {\huge $A$};
\draw[thick,blue,dashed,fill=white]  (25.5,-2)circle (1);
\node[blue] at (25.5,-2) {\huge $B$};

\draw[thick, green] (19,4) -- (29.5,4);
\node[green] at (24.5,5.5) {\huge Environment $\mathsf{e}$};
\node at (24.25,-5.25) {\huge $\Psi$};
\draw[thick]  (22.5,-4.5) rectangle (27.5,-6);

\draw[-triangle 90] [thick]  (25.5,-4.5) to (25.5,-3);
\draw[-triangle 90] [thick]  (23,-4.5) to (23,4);

\draw[-triangle 90] [thick]  (25.5,-1) to (25.5,0.5);
\draw[-triangle 90] [thick]  (25.5,2.5) to (25.5,4);

\node at (25,3) {\huge $\mathsf{a_2}$};
\node at (26,-3.5) {\huge $\mathsf{b_1}$};
\node at (26,-0.5) {\huge $\mathsf{b_2}$};
\node at (25,0) {\huge $\mathsf{a_1}$};
\node at (11.5,4.5) {\huge $\mathsf{e_1}$};
\node at (14,4.5) {\huge $\mathsf{e_2}$};
\node at (39.5,-3.5) {\huge $\mathsf{e_3}$};
\node at (34,4.5) {\huge $\mathsf{e_1}$};
\node at (37.5,4.5) {\huge $\mathsf{e_2}$};
\node at (23,4.5) {\huge $\mathsf{e_1}$};
\node at (25.5,4.5) {\huge $\mathsf{e_2}$};
\draw[-triangle 90] [thick] (38.5,-4.5) -- (39,-3);
\node at (28,-3.5) {\huge $\mathsf{e_3}$};
\draw[-triangle 90] [thick] (27,-4.5) -- (27.5,-3);
\node at (16.5,-3.5) {\huge $\mathsf{e_3}$};
\draw[-triangle 90] [thick] (15.5,-4.5) -- (16,-3);
\end{tikzpicture}
}
    
    \caption{The three causal relations. System $c$ not accessible to $A$ and $B$ is not drawn but is supposed to lie somewhere within the environment $e$.}
    \label{fig:3cr}
\end{figure}
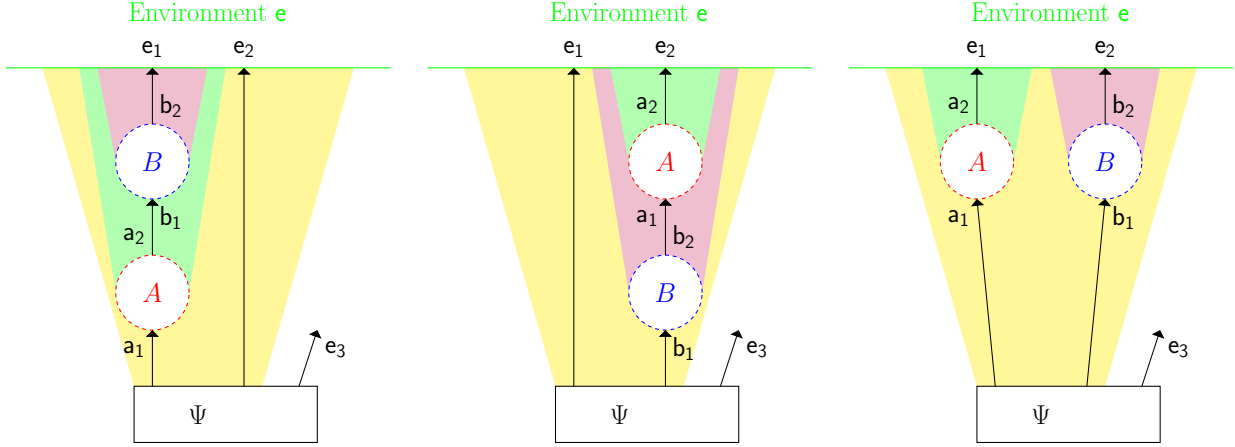

To obtain a concrete model, one needs to fix $\ket{w(\alpha)}^{GABE}$. The $n=3$ (the number of basis vectors for $G$) \textit{harmonic clean models} are some simple models that capture the essential features of causal fluctuations. They describe the situation depicted in Figure \ref{fig:3cr} with the following vector\footnote{Although we use this process matrix to model quantum causal fluctuation, the same mathematical object can describe a mass distribution in positional superposition that induces indefinite causal structure for $A$ and $B$, with $G$ describing the mass degree of freedom \cite{feix2017quantum}.}:
\begin{align}
\ket{w(\alpha)}^{GABE}:=&\alpha_1\ket{1}^g\ket{\Psi(\alpha)}^{a_1 e_2e_3}\ket{I}^{a_2 b_1}\ket{I}^{b_2 e_1}+\alpha_2\ket{2}^g\ket{\Psi(\alpha)}^{e_1 b_1e_3}\ket{I}^{b_2 a_1}\ket{I}^{a_2 e_2}\nonumber
\\
&+\alpha_3\ket{3}^g\ket{\Psi(\alpha)}^{a_1b_1e_3}\ket{I}^{a_2 e_1}\ket{I}^{b_2 e_2}.\label{eq:3crps}
\end{align}
The tripartite state $\ket{\Psi(\alpha)}$ (at the bottom of the pictures in Figure \ref{fig:3cr}) is the ``initial state'' $A$ and $B$ can receive information from. The vector $\ket{I}^{xy}$ (a pure maximally entangled state) is the Choi state representation of the identity channel from $x$ to $y$. In a more general model, both the channels between $A$ and $B$ and the state $\Psi$ may depend on $\alpha$. In the model we analyze we make the assumption that the channels do not. This simplification allows us to find exact answers for the causality measures later on, but a more general study may remove the assumption. The three pictures of Figure \ref{fig:3cr} each depict a definite causal structure between $A$ and $B$, and they correspond to the three terms of (\ref{eq:3crps}). The pictures explain why the systems are coupled through $\ket{\Psi}$ or $\ket{I}$ in the the particular way in (\ref{eq:3crps}).

From the local perspective of $A$ and $B$, the reduced process matrix on $A$ and $B$ takes the form
\begin{align}\label{eq:wcm}
W^{AB}(\alpha)=&\Tr_{GE}W^{GABE}(\alpha)=\sum_{i=1}^3 p_i W_i^{AB}(\alpha),
\\
\rho^{xye_3}=&\ketbra{\Psi(\alpha)}{\Psi(\alpha)}^{xye_3},
\\
\rho^{a_1}=&\rho^x, ~ \rho^{b_1}=\rho^y, ~ \rho^{a_1b_1}=\rho^{xy},\label{eq:rsc}
\\
W_1^{AB}(\alpha):=&\rho^{x}(\alpha)\otimes \Upphi^{a_2b_1}\otimes \pi^{b_2}=\rho^{a_1}(\alpha)\otimes \Upphi^{a_2b_1}\otimes \pi^{b_2},\label{eq:w1}
\\
W_2^{AB}(\alpha):=&\rho^{y}(\alpha)\otimes \Upphi^{a_1b_2}\otimes \pi^{a_2}=\rho^{b_1}(\alpha)\otimes \Upphi^{a_1b_2}\otimes \pi^{a_2},\label{eq:w2}
\\
W_3^{AB}(\alpha):=&\rho^{xy}(\alpha)\otimes \pi^{a_2}\otimes \pi^{b_2}=\rho^{a_1b_1}(\alpha)\otimes \pi^{a_2}\otimes \pi^{b_2},\label{eq:w3}
\end{align}
where $\pi$ is the maximally mixed state. The $\alpha$ dependence of $W_i$ comes from the $\alpha$ dependence of $\Psi$. The meaning of (\ref{eq:rsc}) is that $\rho^{a_1}$ is the reduced state of $\rho^{xye_3}$ on the first subsystem, $\rho^{b_1}$ is the reduced state of $\rho^{xye_3}$ on the second subsystem, and $\rho^{a_1b_1}$ is the reduced state of $\rho^{xye_3}$ on the first and second subsystems.

\subsection{General models}\label{subsec:gm}

As mentioned, one reason the clean models are ``clean'' is the existence of a separate quantum gravitational party $G$. A general model of causal fluctuations may or may not have this property. An example not exhibiting this property was studied in \cite{jia2017quantum} (based on an earlier model in the context of inferring causal structure \cite{maclean2017quantum}). It puts the relations $A\rightarrow B$ and $A-B$ into a coherent mixture through the use of a ``partial swap unitary''. This model is presented just to show that there are reasonable general models that are not clean models, but it will not show up again in subsequent sections.

For $0\le p\le 1$, define the \textit{partial swap channel} $P(p):L(\mathcal{H}^{a}\otimes \mathcal{H}^{a})\rightarrow L(\mathcal{H}^{a}\otimes \mathcal{H}^{a})$ on two subsystems of equal dimension to be the channel induced by the \textit{partial swap unitary}:
\begin{align}
&V(p)=\sqrt{1-p}~\id+\sqrt{p}~ i ~ U_{\text{SW}},
\\
&P(p): \rho\mapsto V(p)\rho V(p)^\dagger.
\end{align}
Here $\id$ is the identity operator and $U_{\text{SW}}:\ket{ij}\mapsto \ket{ji}$ is the swap operator. Define
\begin{align}
W^{a_1b_1}_{a_2}(p):=\Tr_e P_{a_1'a_2'}^{b_1 e}(p)\rho^{a_1 a_1'} N^{a_2'}_{a_2},\label{eq:2crgp}
\end{align}
where $a_1'$ and $a_2'$ are copies of $a_1$ and $a_2$, $\rho$ is an initial state, and $N$ is a channel.

The $\id$ part of $P$ transmits $a_1'$ to $b_1$ and $a_2'$ to the environment $e$, while the $i ~ U_{SW}$ part of $P$ does the opposite, sending $a_2'$ to $b_1$ and $a_1'$ to $e$. The whole setting puts $\id$ and $i ~ U_{SW}$ parts into a ``coherent superposition'', such that $A$ and $B$ partially share a channel from $a_2$ to $b_1$, and partially share a bipartite state on $a_1$ and $b_1$. The output $b_2$ of $B$ is not correlated with $a_1, a_2$ or $b_1$. Its information goes directly to the environment through a unitary channel such that the reduced Choi state on $b_2$ is maximally mixed. The bipartite process matrix on $a_1a_2b_1b_2$ is
\begin{align}\label{eq:gmcs}
W^{AB}(p)=W^{a_1b_1}_{a_2}(p)\otimes W_{b_2}, 
\end{align}
where $W_{b_2}=\pi$ is the maximally mixed state that functions to trace out whatever sent into $b_2$.

\section{Massless and massive sectors}\label{sec:sec}

In classical spacetime, free massless objects travel on the light-cone and free massive particles objects travel within the light-cone. Viewed from an informational perspective, this says that if two parties localized at two events in classical spacetime are restricted to using freely propagating objects to communicate, with massless objects they can only communicate if they are light-like separated, and with massive objects they can communicate only if they are time-like separated. Therefore there is a need to distinguish between massless and massive information carriers.

Although more realistically the parties that send and receive information conduct operations in extended regions rather than at point-like events, it still holds that massless and massive information carriers behave differently such that there is a need to distinguish the massless and massive parts. This distinction between massless and massive objects has no reason to disappear when we move from classical to quantum spacetime, so the models of quantum spacetime presented in previous sections need to account for the difference. We introduce ``massless sector'' and ``massive sector'' in the process matrix models of quantum spacetime in this section.



Suppose the sender uses both a massless object and a massive object in an attempt to transmit information. The state of the information carrier is described by a tensor product of two states, one for the massless object on the massless Hilbert space and one for the massive object on the massive Hilbert space. In general, the massless and massive Hilbert spaces may contain one or more particles, but there will always be a factorization into the massless Hilbert space and the massive Hilbert space. For process matrix model descriptions of quantum spacetime, each system (either input or output) with the Hilbert space $\mathcal{H}$ will factor into a tensor product of a massless sector and a massive sector as $\mathcal{H}=\mathcal{H}^{\circ}\otimes \mathcal{H}^{\bullet}$, where a circle $\circ$ ``not filled with mass'' represents the massless sector, and a circle $\bullet$ ``filled with mass'' represents the massive sector. For example, $A$'s input space $\mathcal{H}^{a_1}$ now becomes $\mathcal{H}^{\asi}\otimes \mathcal{H}^{\avi}$. At the level of systems, the factorization is denoted in the form $a_1a_2b_1b_2=\asi\avi\aso\avo\bsi\bvi\bso\bvo$. We write $A=A^\circ A^\bullet$ so that $A^\circ$ has $\asi\aso$ and $A^\bullet$ has $\avi\avo$, and similarly for $B$. Next we consider how the introduction of sectors impact the process matrices.

\subsection{Clean models}\label{subsec:cm}


As mentioned at the beginning of the section, the causal relations in the massless and massive sectors do not have to agree. Two parties may be causally connected by massless information carriers but not by massive information carriers and vice versa. Each sector has three possible causal relations, so in principle the two sectors have $3^2=9$ causal relations in total. In a clean model each of the nine relations is coupled to a particular state in the quantum gravity party $G$. We denote these states as $\ket{ij}^{\gs\gv}$, where $i=1,2,3$ labels the causal relation in the massless sector, and $j=1,2,3$ labels the causal relation in the massive sector. For example, $\ket{13}^{\gs\gv}$ is coupled to the causal relation $A\rightarrow B$ in the massless sector and $A-B$ in the massive sector.

A bipartite clean model with massless and massive sectors takes the form (the process matrix is induced by the following vector)
\begin{align}\label{eq:cm2s}
\ket{w(\alpha^{\circ\bullet})}^{GABE}=\sum_{i,j=1}^3 \alpha_{ij}^{\circ\bullet}\ket{ij}^{\gs\gv}\ket{w_{ij}}^{ABE}, \text{ with } \alpha_{12}^{\circ\bullet}=\alpha_{21}^{\circ\bullet}=0,
\end{align}
where $\alpha^{\circ\bullet}$ is a complex-vector with the nine probability amplitudes $\alpha_{ij}^{\circ\bullet}$ as components so that $\sum_{i,j=1}^3\abs{\alpha_{ij}^{\circ\bullet}}^2=1$, and $\ket{w_{ij}}^{ABE}$ induce process matrices $W_{ij}^{ABE}=\ketbra{w_{ij}}^{ABE}$ with the causal relation $i$ for the massless sector and $j$ for the massive sector. 

Importantly, the extra conditions $\alpha_{12}^{\circ\bullet}=0$ and $\alpha_{21}^{\circ\bullet}=0$ say that only seven out of the nine logically possible causal relations are present. Mathematically, this is a consequence of the normalization condition of the process matrix framework. It can be proved (e.g., using the Hilbert-Schmidt basis decomposition method of \cite{oreshkov2012quantum}) that if probabilities sum up to one, $\alpha_{12}^{\circ\bullet}=\alpha_{21}^{\circ\bullet}=0$ must hold. Intuitive, there is an informal explanation that the ruled out terms would lead to correlations of the closed-causal-loop type \cite{oreshkov2012quantum}. For example, $\ket{w_{12}}^{ABE}$ may be regarded as a term that allows $A$ to signal to $B$ through the massless sector, and $B$ to signal back to $A$ through the massive sector.

A process matrix induced by (\ref{eq:cm2s}) allows the massless and massive objects to interact, as $\ket{w_{ij}}$ does not factor into the two sectors. Physically, this can describe the interaction of massless and massive objects in quantum spacetime. The following simpler model allows no interaction between the two sectors:
\begin{align}
\ket{w(\alpha^{\circ\bullet})}^{GABE}=\sum_{i,j=1}^3 \alpha_{ij}^{\circ\bullet}\ket{ij}^{\gs\gv}\ket{w_{i}}^{A_\circ B_\circ E_\circ} \ket{w_{j}}^{A_\bullet B_\bullet E_\bullet}, \text{ with } \alpha_{12}^{\circ\bullet}=\alpha_{21}^{\circ\bullet}=0.
\end{align}
Here $\ket{w_{i}}^{\As\Bs\Es}$ and $\ket{w_{j}}^{\Av\Bv\Ev}$ induce process matrices $W_{i}^{\As\Bs\Es}=\ketbra{w_{i}}^{\As\Bs\Es}$ of the massless sector compatible with the causal relation $i$ and  $W_{j}^{\Av\Bv\Ev}=\ketbra{w_{j}}^{\Av\Bv\Ev}$ of the massive sector compatible with the causal relation $j$. Such models, applicable when there is no inter-sector interaction or when the interaction can be ignored, are often simpler to analyze than the general models (\ref{eq:cm2s}). 

With the quantum gravitational party $G$ and the environmental party $E$ traced out, one gets
\begin{align}
W^{AB}=\sum_{i,j=1}^3 p_{ij}^{\circ\bullet} W_i^{A_\circ B_\circ} \otimes W_j^{A_\bullet B_\bullet}, \text{ with } p_{12}^{\circ\bullet}=p_{21}^{\circ\bullet}=0,\label{eq:ctmab}
\end{align}
where $p_{ij}^{\circ\bullet}:=\abs{\alpha_{ij}^{\circ\bullet}}^2$ and $\sum_{i,j=1}^3 p_{ij}^{\circ\bullet}=1$. The process matrices in the individual sectors are obtained by tracing out the other sector
\begin{align}
W^{A_\circ B_\circ}=&\Tr_\bullet W^{AB}=\sum_{i=1}^3 p_{i}^{\circ} W_i^{A_\circ B_\circ}, \quad p_i^{\circ}=\sum_{j=1}^3 p_{ij}^{\circ\bullet}, \quad \sum_{i=1}^3 p_i^{\circ}=1,\label{eq:cmrs}
\\
W^{A_\bullet B_\bullet}=&\Tr_\circ W^{AB}=\sum_{j=1}^3 p_{j}^{\bullet} W_j^{A_\bullet B_\bullet}, \quad p_j^{\bullet}=\sum_{i=1}^3 p_{ij}^{\circ\bullet}, \quad \sum_{j=1}^3 p_j^{\bullet}=1.\label{eq:cmrv}
\end{align}
These are of form of the harmonic clean models (\ref{eq:wcm}). 

\subsection{General models}

A general model does not have the detailed structures of a clean model, and there is less to say about it. We know that on general the normalization condition of the framework automatically implies that in all Hilbert-Schmidt terms of the process matrix the massless and massive sectors cannot signal in different directions. A rather trivial example of a general model is obtained by assigning one partial-swap induced process matrix of the form (\ref{eq:gmcs}) to each sector and taking the tensor product. 

\section{The tendency postulate}\label{sec:tp}



Recall that the tendency postulate says \textit{in quantum spacetime free massless objects tend to trace large quantum causal fluctuations}. The meaning and motivation for the postulate are explained in Section \ref{sec:i}. In this section we give some further discussion on its validity and usefulness, and then focus on implementing the postulate in process matrix models of quantum spacetime.

The tendency postulate is a ``postulate'' motivated by theoretical considerations and at present we do not have enough empirical evidence to confirm or invalidate it. It is an interesting open question what experiments the postulate suggests to test itself. As mentioned in the introduction, the tendency postulate is based on the expectation that quantum gravitational fluctuations are large at small invariant distances. That quantum gravitational fluctuation takes place according to some invariant distance rather than spatial or temporal distance in some preferred reference frame is a major assumption behind the tendency postulate. Models of quantum gravity that break this assumption and the induced phenomenology have been extensively studied \cite{mattingly2005modern, amelino2013quantum}. Currently data seem to be in favour of the assumption. Unless new data suggest otherwise, we take the assumption to hold and regard the tendency postulate as reasonable. 

Regarding the validity of the tendency postulate, another point worth discussing is the Jonsson,  Mart{\'i}n-Mart{\'i}nez and Kempf (JM-MK) effect of subluminal information propagation mediated by massless quantum fields \cite{jonsson2015information}. In 3+1 classical spacetime with non-trivial curvature massless fields do not have to propagate on the light-cone (violation of the Strong Huygens Principle). The JM-MK effect exploits this feature to show that using Unruh-DeWitt detectors time-like separated parties can transmit information with massless quantum fields. The JM-MK effect does not contradict the tendency postulate, which is based on the intuition that massless objects in classical spacetime travels at the speed of light. Although information can be transmitted in the sense that the channel capacity is positive, energy does not propagate from the sender to the receiver (Instead, the receiver has to supply a signal-dependent amount of work to collect the information). Because ``massless objects'' in the tendency postulate denote objects that carry energy, there is not contradiction with the JM-MK effect. In addition, the tendency postulate speaks about the \textit{tendency} of massless objects propagation, and is based on the expectation that they tend (with large probability) to travel between regions with small absolute invariant distance separation. This expectation does not contradict the JM-MK effect, which takes place with small probability. 

We mentioned in the Introduction that the tendency postulate is useful in developing other concepts for quantum spacetime that were based on null geodesics for classical spacetime, in judging whether a model of quantum spacetime is typical, and guiding the construction of concrete models of quantum spacetime. Here we explain further what we mean by ``judging'' for the second use. The tendency postulate says that ``free massless objects \textit{tend} to trace...'', rather than ``free massless objects trace...''. This is because information and energy propagation in quantum spacetime is stochastic. Due to the intrinsic randomness of quantum spacetime there is no guarantee that massless objects always trace large causal fluctuation. Instead, one can only expect that there is a large chance this happens -- hence the word ``tend''. If a model fits this tendency described by the tendency postulate, we say that it is typical. The tendency postulate does not offer a necessary condition for a model to be valid. An atypical model may with small chance arise in nature. The tendency postulate does not offer a sufficient condition for a model to be valid either. For example, at the level of processes $W^{AB}$ reduced to $A$ and $B$, a harmonic clean model (\ref{eq:ctmab}) is not distinguishable from a classical mixture process in terms of mathematical description. It is the existence of a correlated quantum gravitational system $G$ that makes the harmonic clean models $W^{ABG}$ describe coherent quantum indefinite structure. Yet the tendency postulate conditions to be presented below apply to the reduced processes $W^{AB}$, so some classical mixture processes not describing quantum spacetime may pass the typicality test of the tendency postulate. Although the tendency postulate does not have much exclusive power for individual models, it is useful as a statistical criterion. If a theory assigns an ensemble of models and most of them do not obey the tendency postulate conditions, then the theory may be ruled out with large confidence. Again, the tendency postulate is analogous to the second law of thermodynamics in these respects.

For the presentations in the subsequent subsections, the picture to bear in mind is that we are picking pairs of highly localized parties in quantum spacetime and examine their process matrices to see if massless objects tend to propagate between these two parties. By considering highly localized parties we exclude too much extended parties between which massless objects can trivially propagate. It is also assumed that the parties do not overlap.

To be precise, we will consider quantum spacetime models for collaborative parties in the examples of the following subsections. A general model of quantum spacetime should account for all the possible choices of the parties including collaborative parties who make conducive choices for transmitting information and adversarial parties who do not. For example, one choice of the sending party at the disadvantage of information transmission is to send signals to the directions opposite to the receiving party. A general quantum spacetime model must include a part that describes what the other party receives in this case. This possibility can be accounted for, for example, using completely depolarizing channels. However, the adversarial cases are not really relevant for implementing the tendency postulate (If one does not send massless objects to some direction, massless objects will trivially not tend to trace that direction.). Therefore in the following we focus on the relevant parts which describe helpful resources for information transmission quantum spacetime offers when the parties try their best to communicate in both the massless and the massive sectors.

\subsection{Clean models}

For clean models, the tendency of massless objects to trace between $A$ and $B$ is measured by $p^\circ_{\text{connect}}=p_1^\circ+p_2^\circ$, which accounts for causal connectedness in both directions. The larger $p^\circ_{\text{connect}}$ is, the greater the tendency is. The tendency postulate states that $p^\circ_{\text{connect}}$ tends to large when the quantum gravitational fluctuation is large for $A$ and $B$.

Because $p_1,p_2,p_3$ represent the squared amplitudes for the three causal relations, a large quantum causal fluctuation means that they are comparable, i.e., $p_1\approx p_2 \approx p_3$. In the massive sector (\ref{eq:cmrs}), this means that $p_1^\bullet,p_2^\bullet,p_3^\bullet$ are comparable. If we interpret the tendency postulate as saying that ``in quantum spacetime free massless objects tend to trace large quantum causal fluctuation in the massive sector'', then we have that for typical models,
\begin{align}\label{eq:tpctv}
p^\circ_{\text{connect}} \text{ is large if and only if }p_1^\bullet\approx p_2^\bullet \approx p_3^\bullet.
\end{align}

Note that (\ref{eq:tpctv}) is an if and only if condition. This condition in fact interprets the tendency postulate as saying that ``in quantum spacetime if the causal quantum causal fluctuation in the massive sector for a pair of parties is large, then free massless objects must tend trace this pair of parties'', and in addition that ``in quantum spacetime if free massless objects tend to trace a pair of parties then the causal quantum causal fluctuation in the massive sector for this pair of parties must be large''. Is it possible to interpret the tendency postulate as stating only one of the if and the only if conditions? Recall from Section \ref{sec:i} the motivation for the tendency postulate, that small absolute values of proper distance accompany large quantum gravitational fluctuations. This companionship is two-way, so the tendency postulate should be interpreted as implying both the if and the only if conditions.


An example of a clean model that is typical under condition (\ref{eq:tpctv}) comes with
\begin{align}\label{eq:tm}
p_{ij}^{\circ\bullet}=
\begin{pmatrix}
    1/3-1/10^{10} & 0 & 1/6-1/10^{10} \\
    0 & 1/3-1/10^{10} & 1/6-1/10^{10} \\
    1/10^{10} & 1/10^{10} & 2/10^{10} \\
\end{pmatrix},
\end{align}
where $i$ labels the row and $j$ labels the column. Recall from (\ref{eq:cmrs}) and (\ref{eq:cmrv}) that $p_i^{\circ}=\sum_{j}p_{ij}^{\circ\bullet}$ and $p_j^{\bullet}=\sum_{i}p_{ij}^{\circ\bullet}$. It is easy to see that $p_1^\circ= p_2^\circ =1/2-2/10^{10}$, $p^\circ_{\text{connect}}=p_1^\circ+ p_2^\circ=1-4/10^{10}\approx 1$, and $p_1^\bullet= p_2^\bullet = p_3^\bullet=1/3$. In addition, the consistency conditions $p_{12}^{\circ\bullet}=p_{21}^{\circ\bullet}=0$ in (\ref{eq:ctmab}) are satisfied. An example of a clean model that is not typical under condition (\ref{eq:tpctv}) comes with
\begin{align}\label{eq:atm}
p_{ij}^{\circ\bullet}=
\begin{pmatrix}
    1/10^{10} & 0 & 1/2-3/10^{10} \\
    0 & 1/10^{10} & 1/2-3/10^{10} \\
    1/10^{10} & 1/10^{10} & 2/10^{10} \\
\end{pmatrix}.
\end{align}
Here $p_1^\circ= p_2^\circ =1/2-2/10^{10}$, $p^\circ_{\text{connect}}=p_1^\circ+ p_2^\circ=1-4/10^{10}\approx 1$, which means that massless objects tend to propagate between $A$ and $B$. However, $p_1^\bullet= p_2^\bullet =2/10^{10}, p_3^\bullet=1-2/10^{10}$, which means that the causal fluctuation in the massive sector is very small.

In condition (\ref{eq:tpctv}), the causal fluctuation in the massive sector determines if $p^\circ_{\text{connect}}$ tend to be large regardless of whether the causal fluctuation in the massless sector is large. One might ask in addition that causal fluctuation in the massless sector itself has to be large in order for $p^\circ_{\text{connect}}$ to tend to be large. This would mean that for typical models
\begin{align}\label{eq:tpctvs}
p^\circ_{\text{connect}} \text{ is large if and only if }p_1^\bullet\approx p_2^\bullet \approx p_3^\bullet\text{  and }p_1^\circ\approx p_2^\circ \approx p_3^\circ.
\end{align}
This is a much more restrictive constraint. Because $p_1^\circ\approx p_2^\circ \approx p_3^\circ$ has to be met in order that $p^\circ_{\text{connect}}$ tend to be large, the largest $p^\circ_{\text{connect}}$ that obeys the tendency is about $2/3$. Hence any $p^\circ_3\lesssim 1/3$ would not obey the tendency.

Furthermore, there is the logical possibility that for typical models
\begin{align}\label{eq:tpcts}
p^\circ_{\text{connect}} \text{ is large if and only if }p_1^\circ\approx p_2^\circ \approx p_3^\circ.
\end{align}
This would mean that regardless of what happens in the massive sector, large causal fluctuations in the massless sector alone implies that massless objects tend to propagate between such parties, and a large $p^\circ_{\text{connect}}$ does not imply anything about the massive sector at all. Ever further, there is the possibility that for typical models
\begin{align}\label{eq:tpctvors}
p^\circ_{\text{connect}} \text{ is large if and only if }p_1^\bullet\approx p_2^\bullet \approx p_3^\bullet\text{  or }p_1^\circ\approx p_2^\circ \approx p_3^\circ.
\end{align}
This would mean that large causal fluctuation in either sector corresponds to massless objects tending to propagate between such parties.

All of (\ref{eq:tpctv}), (\ref{eq:tpctvs}), (\ref{eq:tpcts}) and (\ref{eq:tpctvors}) are logically possible consequences of the tendency postulate. For (\ref{eq:tpctv}) we presented a typical example (\ref{eq:tm}) and an atypical example (\ref{eq:atm}) to show that (\ref{eq:tpctv}) is a non-trivial condition. It is not hard to see that (\ref{eq:tpctvs}), (\ref{eq:tpcts}) and (\ref{eq:tpctvors}) are not-trivial either, so in principle all four conditions are meaningful candidates. In addition, these conditions only characterize typical models at a qualitative level without specifying quantitatively what ``large'' means. The questions regarding the choice among (\ref{eq:tpctv}), (\ref{eq:tpctvs}), (\ref{eq:tpcts}) and (\ref{eq:tpctvors}), and regarding a quantitative specification of how large is ``large'' cannot be addressed with the current statement of the tendency postulate. Some additional inputs, theoretical or phenomenological, are needed to decide among the possibilities. We leave the task of addressing these two questions to future work.


\subsection{General models}

In general models (e.g. (\ref{eq:gmcs})) the amplitudes $\alpha_{ij}^{\circ\bullet}$ and the derived quantities $p_i^\circ$ and $p_j^\bullet$ may not exist, so the considerations in the previous subsection do not generalize. It was possible to implement the tendency postulate based on the quantities $p_i^\circ$ and $p_j^\bullet$ because they characterize strengths for different causal relations. For general models, one can instead use \textit{causality measures} to quantify the strength of causal relations.

Causality measures, like entanglement measures, refer to a family of functions. In short, a bipartite causality measure $\mu^\rightarrow(G^{AB})$ is a function on bipartite correlations $G^{AB}$ shared by $A$ and $B$ (such as the process matrices $W^{AB}$) and measures its causal strength in the $A\rightarrow B$ direction. We list the defining axioms in Appendix \ref{app:oc} and refer the readers to \cite{jia2018quantifying} for a more detailed study of these functions.

The idea to implement the tendency postulate is to use clean models which we know are typical according to (\ref{eq:tpctv}), (\ref{eq:tpctvs}), (\ref{eq:tpcts}) or (\ref{eq:tpctvors}) to calibrate typical general models. If a general model is close enough to a typical clean model, then we regard it as typical as well. Whether or not two models are close is determined by some criterion based on causality measures. For example, one criterion could be that if the process matrices $G_1$ and $G_2$ obey $\abs{\mu^\rightarrow(G_1)-\mu^\rightarrow(G_2)}\le d^\rightarrow$ and $\abs{\mu^\leftarrow(G_1)-\mu^\leftarrow(G_2)}\le d^\leftarrow$ for some threshold values $d^\rightarrow$ and $d^\leftarrow$, then $G_1$ and $G_2$ are close. In general there is some flexibility (e.g., use different numbers causality measures) in choosing the criterion for the closeness of models in terms of causal strength, just like there is some flexibility in choosing the criterion for states to contain similar amounts of entanglement. Different contexts suggest different criteria. Appendix \ref{app:ccm} exemplifies the above ideas using causality measures based on one-shot quantum communication capacities.

\section{Black hole leakage}\label{sec:ss}

The tendency postulate suggest that black holes in quantum spacetime allows information leakage without violating locality. In this section we sketch this suggestion. At face value the suggestion seems to be an obvious contradiction, because by definition a black hole contains a causal boundary of information propagation, which if surpassed would result in a violation of locality. We argue below that the concepts of black hole, causal boundary, and locality should be upgraded for quantum spacetime. After this is done, black holes allowing information leakage without violating locality is nothing contradictory.

\subsection{No superluminal signalling}

In the present context we take ``locality'' to mean ``no superluminal signalling''. The tendency postulate is based on the expectation that quantum causal fluctuations tend to be large when the relevant parties are close. When the separation is small in comparison to the Planck scale causal fluctuations are expected to be very significant. On the other end, when the separation is very large although causal fluctuations are expected to be very small, one cannot expect them to vanish. The situation is analogous to quantum fluctuations of matter fields, which is significant at small scales, but generically not exactly vanishing at large scales. 

The expected generically present quantum causal fluctuations at all scales combined with the tendency postulate seem to imply ``superluminal signalling''. If one takes what the tendency postulate indicates as where massless objects tend to travel as the light-cone, then generic causal fluctuations imply some finite chances that massive or massless objects travel beyond the light-cone. For example, consider a harmonic clean model associated with $A$ and $B$ with $p_3^\bullet\gg p_1^\bullet,p_2^\bullet$ that is typical according to (\ref{eq:tpctv}). For this pair of parties $p^\circ_\text{connect}$ is small and massless objects will not tend to travel between them. In terms of classical intuition, $A$ is predominantly spacelike separated from $B$, as opposed to the case of $p_3^\bullet\ll p_1^\bullet$ or $p_3^\bullet\ll p_2^\bullet$, which in classical intuition corresponds to $A$ being predominantly causally connected to $B$. One might then take $A$ and $B$ to lie outside of each other's light-cone. Yet if the quantum gravitational system is measured to be in $\ket{1}$ and $\ket{2}$ there are some finite chances $p_1^\bullet$ and $p_2^\bullet$ that even massive objects can mediate between the parties. In addition, for any $p_1>0$ or $p_2>0$ there is positive quantum communication capacity from one direction or the other as shown in Appendix \ref{app:oc}. Do these imply superluminal signalling?

The answer is no. First of all superluminal signalling as usually conceived is a concept based on classical spacetime. In quantum spacetime its meaning needs to be specified. We take \textbf{superluminal signalling} in both classical and quantum spacetime to mean that \textit{there exists a pair of parties which are causally connected by some non-light objects but are not causally connected by light}. Importantly, we allow reflections along the way of the propagation. This makes sure that light is not beaten by non-light objects simply because non-light objects travel ``slower''. For example, in classical spacetime freely propagating massive objects can reach places inside the light-cone where freely propagating massless objects cannot reach. This certainly does not imply that these massive objects are ``superluminal''. We allow reflections (such as using mirrors) along the way of the propagation in order for massless objects to turn their propagation direction so they can reach places inside the light-cone. In classical spacetime ensures that light can reach anywhere within the light-cone but nowhere outside of the light-cone, so the above interpretation of the meaning of superluminal signalling works. We assume that it works for quantum spacetime as well. In both quantum and classical spacetime, if some causal connection through non-light objects cannot be matched by light even through interventions by reflection one must conclude that there is superluminal signalling. 

Coming back to the original question whether the tendency postulate implies superluminal signalling, we reach the negative conclusion by simply noting that for any pair of parties, generic causal fluctuations will be present for both the light and non-light sectors. If there is a finite probability for non-light objects to causally connect a pair of parties, there is also a finite probability for light to do so (excluding the ``measure zero'' case that the probability for causal connectedness by light is precisely zero). Therefore non-light objects are never unmatched by light, and there is no superluminal signalling.

\subsection{Quantum-classical correspondence}

In the previous subsection we argued that the models we studied do not allow superluminal signalling, if the meaning of ``superluminal signalling'' is expressed in a way that applies to quantum spacetime. An essential feature about the models of quantum spacetime we studied is that they are probabilistic, which sharply contrasts standard descriptions of classical spacetime which are deterministic. Applying the intuitions about the deterministic classical spacetime indiscriminately to the probabilistic quantum spacetime would result in confusions, such as the suspicion of violation of causality we discussed in the previous subsection. Conversely, some features and effects about quantum spacetime, such as the black hole information leakage we will discuss in the next subsection, would appear counter-intuitive to the someone who is used to thinking in terms of classical spacetime. An analogy can be drawn with the various ``paradoxes'' about quantum theory when one exerts classical intuitions. Standard quantum theory is local and obeys the ``no-signalling principle''. However, if one tries to reproduce the statistics of quantum theory using certain classical models Bell's theorem says that there must be non-locality \cite{bell1964instein}. Similarly, standard quantum theory itself is non-contextual.\footnote{In the sense of Spekken's definition of contextuality \cite{spekkens2005contextuality}.} However, if one tries to reproduce its statistics using certain classical models Bell-Kochen-Specker theorem says that there must be contextuality \cite{bell1966problem, kochen1967problem}. 

Suppose one wants to generalize the originally classical notion of causal boundary to quantum spacetime. In a classical spacetime modelled as a differentiable manifold, boundary is a clearly specified concept. In quantum spacetime it is less clear what a boundary is. The type of models of quantum spacetime we study refers to a set of parties in quantum spacetime, and these parties do not form a continuum. Whatever the concept of causal boundary will be in quantum spacetime, a reasonable expectation is that it separates the parties into those that are causally connected to (or more specifically causally precede or succeed) certain parties and those that are not causally connected to these certain parties. Even this rather weak expectation cannot be met. Generically, any two parties $A$ and $B$ will experience causal fluctuations that allows signalling. For instance, the harmonic toy models (\ref{eq:wcm}) associated with $A$ and $B$ generically have non-zero $\alpha_i$ for all of $i=1,2,3$. These will induce signalling in both the $A$ to $B$ and the $B$ to $A$ directions. Because of this generic causal connectedness, one cannot set up a non-trivial causal boundary in quantum spacetime that meets the expectation of ``separating parties'' mentioned above.

Because many causal-fluctuation-induced causal connections are very weak, one could coarse-grain out those very weak causal connections to obtain a non-trivial causal separation among all the parties. This could be achieved by qualifying causal connectedness through causality measures such as the one-shot quantum capacity (Appendix \ref{app:oc}). One requires that there is causal connection in some direction only if the causal strength in this direction as measured by some causality measures is above certain thresholds. This method can be expected to generate a meaningful quantum-classical correspondence of spacetime. If the tendency postulate holds, then after very weak causal connections are removed two parties are causally connected by free massless objects only if their causal fluctuation used to be large. This kind of free massless object causal connection can be used to set up the classical light-cones in a quantum-classical spacetime correspondence. However, it is possible that due to the original stochasticity of quantum spacetime some pairs of parties in the corresponding classical spacetime still have strong causal connections even if they are separated by the classical light-cone as a causal boundary. Such an apparent violation of locality would be an example of a paradox that arises when one attempts to put on classical notions such as strict causal boundaries to quantum spacetime.

Another way definite causal relations among the parties emerge is through measurements of the quantum gravitational degrees of freedom. For instance in the harmonic clean model (\ref{eq:wcm}), measuring $G$ in the basis vectors will reduce the bipartite process matrix to one of $W^{AB}_i$, which has a definite causal relation $A\rightarrow B, A-B,$ or $A\leftarrow B$. For a general model the same reduction to a definite causal relation will take place if the measurement is chosen suitably. If all quantum gravitational degrees of freedom are measured the causal relations between any pair of parties will reduce to definite ones, and the whole spacetime will now possess a definite causal structure. Many of those pairs of parties that were causally connected only due to small causal fluctuations will reduce to being causally disconnected. It is now possible to establish a non-trivial causal boundary that meets the separation requirement. 

\subsection{Black holes}

A message to be drawn from the previous discussions is that one should not define ``locality'' based on classical spacetime notions in studies for which the quantum aspects of spacetime are important. In particular, in studying truly quantum black holes\footnote{As opposed to quantum matter in a classical black hole spacetime, which should better be called ``quantum-classical black hole'' rather than ``quantum black hole''.} one should better not use the classical spacetime notion of locality to judge whether a model contains exotic features. 

The analogy with ordinary quantum theory and Bell's theorem mentioned in the previous subsection applies here. Ordinary quantum theory obeys no-signalling and is not exotic in terms of locality when it is considered in itself. It is only a classical model which matches quantum statistics that has to exhibit nonlocality and appear exotic. If the common belief that Sagittarius A* is an astrophysical black hole holds, its apparent horizon in the classical spacetime description likely is placed very near the center of the Galaxy and far away from the Earth. Yet if information from within the object reaches a spacetime party where someone reads this sentence on Earth, it should not be considered extravagant given the models of quantum spacetime that exhibits causal fluctuation we studied above. We see that generically there is a probability amplitude for any pair of parties to be causally related through causal fluctuation. In fact, Theorem \ref{th:osec} shows that there will be positive quantum communication capacities from $A$ to $B$ for any non-vanishing probability amplitude for $A\rightarrow B$. If one models these effects in a classical description of spacetime, and if there is an actual event horizon bounding the black hole and the events on Earth, it would appear that ``locality'' is violated. However, this is analogous to having nonlocality in a classical hidden variable to match quantum statistics. The quantum theory in itself is actually local. For foundational studies classical hidden variable models are valuable because, for instance, they offer candidates to solve the measurement problem. On the other hand, for studying quantum gravity and understanding truly quantum black holes in particular it is preferable to study quantum models of spacetime such as the ones presented above. Then there is no violation of locality.

Along this line, the tendency postulate suggest the following picture of information propagation for truly quantum black holes. Classical notions of horizons generated by null geodesics correspond in quantum spacetime to regions where many pairs of parties exhibit large causal fluctuations among themselves. As argued in the previous subsections, in quantum spacetime these regions in no way offer a strict causal boundary and information can, albeit usually with very small probability, transmit ``across'' such regions. 

What this picture says about the information-loss problem is hard to estimate at present. The source of the problem, Hawking radiation, is really only an effect derived within the semiclassical theory which studies classical spacetime black holes. Even the very existence of the effect is subject to debate \cite{helfer2003black}. The obvious task is to study if the effect is modified in quantum spacetime.

\section{Conclusion}\label{sec:do}

We proposed a tendency postulate that tells where free massless objects tend to travel in quantum spacetime, just as null geodesics do in classical spacetime. We introduced massless and massive sectors for models of quantum spacetime, and implemented the tendency postulate concretely as characterizations of typical models of quantum spacetime in the process matrix framework. The models we studied suggest that black holes in quantum spacetime allow information leakage without violating locality.

At the current stage, the tendency postulate qualitatively characterizes of the tendency of massless object propagation. It is conceivable that it gets improved into a quantitative characterization that specifies more details of what a typical model should obey in the massless and massive sectors. 


The tendency postulate is introduced to upgrade the notion of null geodesic to quantum spacetime. It should pave the way to study quantum spacetime analogs of other classical spacetime notions based on null geodesics, such as the null geodesic congruence, the light-cone, the light-sheet, and the various horizons. 
In terms of quantum black holes, there have been various studies on the possibility that the information loss paradox gets resolved through non-local information propagation. The present paper offers a similar but different picture that non-locality is not needed for information to leak out of black holes. The opportunities to study this picture more concretely with explicit models is open.

\section*{Acknowledgement}
I am very grateful to Lucien Hardy and Achim Kempf for guidance and support as supervisors. The various discussions we had helped to improve the work from its earlier versions significantly. I thank Jason Pye for a valuable discussion on different ways to formulate the tendency postulate. I thank Fabio Costa for the suggestion during the Spacetime and Information 2017 Workshop of using communication capacity as a quantitative measure of causal structure.

Research at Perimeter Institute is supported by the Government of Canada through the Department of Innovation, Science and Economic Development Canada and by the Province of Ontario through the Ministry of Research, Innovation and Science.  This work is supported by a grant from the John Templeton Foundation. The opinions expressed in this work are those of the author's and do not necessarily reflect the views of the John Templeton Foundation.

\newpage
\appendix

\section{Causality measures}\label{app:oc}

We review some results from \cite{jia2018quantifying} about quantitative measures of causality. 

A \textbf{causality measure} $\mu^{A\rightarrow B}(G)$ on parties $A$ and $B$ sharing the correlation $G$ is a real-valued function obeying the following axioms:
\begin{enumerate}
\item $\mu^{A\rightarrow B}(G)$ is non-increasing under local operations within $A$ and $B$.
\item $\mu^{A\rightarrow B}(G) \ge 0$.
\item $\mu^{A\rightarrow B}(G) > 0$ only if $A$ can signal to $B$ using $G$.
\end{enumerate}
Here ``$A$ can signal to $B$ using $G$'' means that by exploiting the correlation $G$, $A$ can change the measurement outcome probabilities of $B$ by choosing different operations. A \textbf{normalized causality measure} further obeys $\sup_G \mu^{A\rightarrow B}(G)=1$ so that $0\le \mu^{A\rightarrow B}(G)\le 1$ for all $G$. The causality measure $\mu^{A\leftarrow B}(G)$ in the opposite direction is defined similarly except that it obeys Axiom 3 with $A$ and $B$ swapped.

Two very important causality measures for quantum models are the one-shot entanglement transmission capacity and the one-shot subspace transmission capacity. They measure the capacities of a correlation in transmitting quantum entangled states and in transmitting pure quantum states when some finite imprecision $\epsilon\in[0,1]$ in the fidelity is allowed (see \cite{jia2018quantifying} for the exact definitions). Let $A$ and $B$ share a correlation $G^{AB}$, which within the process matrix framework would be a process matrix. Depending on the direction of transmission there are the $A$ to $B$ and $B$ to $A$ entanglement transmission capacities $Q_{\text{ent}}^\rightarrow(G^{AB};\epsilon)$ and $Q_{\text{ent}}^\leftarrow(G^{AB};\epsilon)$, and the $A$ to $B$ and $B$ to $A$ subspace transmission capacities $Q_{\text{sub}}^\rightarrow(G^{AB};\epsilon)$ and $Q_{\text{sub}}^\leftarrow(G^{AB};\epsilon)$.

While the one-shot subspace transmission capacities depend on the yet unspecified dependence of $\Psi$ on $\alpha$, the one-shot entanglement transmission capacities can be solved exactly for the harmonic clean models.
\begin{theorem}\label{th:osec}
For $W^{AB}$ in the family of harmonic clean models (\ref{eq:3crps}),
\begin{align}\label{eq:ec}
Q_{\text{ent}}^\rightarrow(W^{AB};\epsilon)=
\begin{cases}
\log\Big(\max\Big\{m\in \mathbb{N}: m\le \sqrt{\frac{1}{1-\frac{\epsilon}{1-p_1}}}\text{ and }m\le \abs{a_2}\Big\}\Big), & p_1<1-\epsilon,
\\
\log\abs{a_2}, & p_1\ge 1-\epsilon.
\end{cases}
\\\label{eq:ecl}
Q_{\text{ent}}^\leftarrow(W^{AB};\epsilon)=
\begin{cases}
\log\Big(\max\Big\{m\in \mathbb{N}: m\le \sqrt{\frac{1}{1-\frac{\epsilon}{1-p_2}}}\text{ and }m\le \abs{b_2}\Big\}\Big), & p_2<1-\epsilon,
\\
\log\abs{b_2}, & p_2\ge 1-\epsilon.
\end{cases}
\end{align}
\end{theorem}
$Q_{\text{ent}}^\rightarrow(W^{AB};\epsilon)$ is a non-decreasing functions of $p_1$, which makes intuitive sense because as the amplitude of $A$ in the causal past of $B$ increase we expect a better capacity. 

For fixed $\abs{a_2}$, it follows directly from (\ref{eq:ec}) that $Q_{\text{ent}}^\rightarrow(W^{AB};\epsilon)$ takes the maximum value $\log\abs{a_2}$ if and only if 
\begin{align}
    \frac{\epsilon}{1-p_1}\ge 1-\frac{1}{\abs{a_2}^2}.
\end{align}
On the other hand, $Q_{\text{ent}}^\rightarrow(W^{AB};\epsilon)=0$ if and only if
\begin{align}\label{eq:ccz}
p_1<1-\frac{4}{3}\epsilon.
\end{align}
For any harmonic clean model $W$ with $p_1>0$, there is always an $\epsilon$ that allows it to have a larger one-shot entanglement transmission capacity than non-signalling correlations.

There is an interesting theorem that says the family of one-shot entanglement transmission capacities can be used to reconstruct the causally relevant part of the unknown harmonic clean model.
\begin{theorem}\label{th:etr}
Let $W^{AB}$ be any unknown harmonic clean model. Then the capacities $Q_{\text{ent}}^\rightarrow(W^{AB};\epsilon)$ and $Q_{\text{ent}}^\leftarrow(W^{AB};\epsilon)$ indexed by $\epsilon$ determines $\abs{\alpha}:=(\abs{\alpha_1},\abs{\alpha_2},\abs{\alpha_3})$. If $\abs{\alpha_3}\neq 1$, they also determine the subsystem dimensions.
\end{theorem}

\section{Characterizing the closeness of models using  one-shot entanglement transmission capacities}\label{app:ccm}

We demonstrate how to use the one-shot entanglement transmission capacities as causality measures to characterize the closeness of models.

Let $Z^{AB}$ and $W^{AB}$ be two process matrices. There are two families of causality measures $Q_{\text{ent}}^\rightarrow(\epsilon)$ and $Q_{\text{ent}}^\leftarrow(\epsilon)$ parametrized by $\epsilon$. In general one can choose an arbitrary subset of these measures to characterize the closeness of $Z^{AB}$ and $W^{AB}$. For example, pick $\epsilon=1/100$ for both directions, set the threshold values $d^\rightarrow=2$ and $d^\leftarrow=3$, and say that $Z^{AB}$ and $W^{AB}$ are close if and only if
\begin{align}
\abs{Q_{\text{ent}}^\rightarrow(Z^{AB};1/100)-Q_{\text{ent}}^\rightarrow(W^{AB};1/100)}\le&d^\rightarrow=2,
\\
\abs{Q_{\text{ent}}^\leftarrow(Z^{AB};1/100)-Q_{\text{ent}}^\leftarrow(W^{AB};1/100)}\le&d^\leftarrow=3.
\end{align}
This gives one criterion for when two process matrices are close.

In general, for $Q_{\text{ent}}^\rightarrow(\epsilon)$ one picks a set of $\epsilon\in I^\rightarrow$ and for $Q_{\text{ent}}^\leftarrow(\epsilon)$ one picks a set of $\epsilon\in I^\leftarrow$ to use. For each $\epsilon\in I^\rightarrow$ we need to choose a threshold $d^\rightarrow_{\epsilon}$ and for each $\epsilon\in I^\leftarrow$ we need to choose a threshold $d^\leftarrow_{\epsilon}$ to qualify closeness so that $Z^{AB}$ and $W^{AB}$ if and only if
\begin{align}
\abs{Q_{\text{ent}}^\rightarrow(Z^{AB};\epsilon)-Q_{\text{ent}}^\rightarrow(W^{AB};\epsilon)}\le&d^\rightarrow_{\epsilon},\quad \epsilon \in I^\rightarrow,
\\
\abs{Q_{\text{ent}}^\leftarrow(Z^{AB};\epsilon)-Q_{\text{ent}}^\leftarrow(W^{AB};\epsilon)}\le&d^\leftarrow_{\epsilon}, \quad \epsilon\in I^\leftarrow.
\end{align}
The choices of $I^\rightarrow$, $I^\leftarrow$, $\{d^\rightarrow_{\epsilon}\}_{\epsilon\in I^\rightarrow}$, and $\{d^\leftarrow_{\epsilon}\}_{\epsilon\in I^\leftarrow}$ will depend on the particular context of application. One reasonable criterion in choosing $\{d^\rightarrow_{\epsilon}\}_{\epsilon\in I^\rightarrow}$ and $\{d^\leftarrow_{\epsilon}\}_{\epsilon\in I^\leftarrow}$ is that they should ``work within harmonic clean models'': Harmonic clean models close to a typical one according to $\{d^\rightarrow_{\epsilon}\}_{\epsilon\in I^\rightarrow}$ and $\{d^\leftarrow_{\epsilon}\}_{\epsilon\in I^\leftarrow}$ should be typical to start with.


The possible conditions (\ref{eq:tpctv}), (\ref{eq:tpctvs}), (\ref{eq:tpcts}) and (\ref{eq:tpctvors}) of the tendency postulate refer to the causal strength of the massive sector, the massless sector, or both. To account for the reference to particular sectors we may specialize the capacities $Q_{\text{ent}}^\rightarrow(W^{AB};\epsilon)$ to particular sectors. For example, (\ref{eq:tpctv}) refers to the massive sector so we can use $Q_{\text{ent}}^\rightarrow(W^{A_\bullet B_\bullet};\epsilon)$ correpondingly, where $W^{A_\bullet B_\bullet}$ is given in (\ref{eq:cmrv}). Another thing to note is that the comparison of capacities between $W^{AB}$ and $Z^{AB}$ to characterize causal strength is only fair when they live on systems of equal dimensions. When one needs to compare correlations on systems with different dimensions, one can use the measure normalization procedure given in \cite{jia2018quantifying}.

\bibliographystyle{apsrev}
\bibliography{bib.bib}

\end{document}